\documentclass{ws-p8-50x6-00}
\usepackage{epsfig}
\def\npb#1#2#3{    {\it Nucl. Phys. }{\bf B #1} (19#2) #3}

\def\plb#1#2#3{    {\it Phys. Lett. }{\bf B #1} (19#2) #3}
\def\prd#1#2#3{    {\it Phys. Rev. }{\bf D #1} (19#2) #3}

\def\prl#1#2#3{    {\it Phys. Rev. Lett. }{\bf #1} (19#2) #3}

\begin{document}

\title{Cosmological implications of Supersymmetric CP violating phases}

\author{Shaaban Khalil}

\address{Departmento de Fisica Te\'orica, C.XI, Universidad
Aut\'onoma de Madrid, 28049 Cantoblanco, Madrid, Spain.\\
\vspace*{2mm} and\\ Ain Shams University, Faculty of Science,
Cairo 11566, Egypt.}


\maketitle

\abstracts{ We show that large SUSY phases have no significant effect on
the relic density of the lightest supersymmetric particle (LSP).
However, they are very significant for the detection rates. We emphasise
that the phase of the trilinear coupling increase the direct and indirect
detection rates.}

In supersymmetric (SUSY) models there are many new CP violating
phases beyond the phase $\delta_{CKM}$ of the
Cabibbo-Kobayashi-Maskawa (CKM) mixing matrix. They arise mainly
from the soft SUSY breaking parameters which are in general
complex.  These phases give large one loop contributions to the
electric dipole moments (EDM) of the neutron and electron which
exceed the current limits. Hence, SUSY phases are generally quite
constrained, they have to be of order $10^{-3}$ for SUSY particle
masses of order 100 GeV. However, it was suggested that there are
internal cancellations among various contribution to the EDM
(including the chromoelectric and purely gluonic operator
contributions) whereby allowing for large CP phases~\cite{nath}.
We have shown that in the effective supergravity derived from
string theory, such cancellation is accidental and it only occurs
at few points in the parameter space~\cite{barr}. Recently, it was
argued that the non universal gaugino masses and their relative
phases are crucial for having sufficient cancellations among the
contributions to EDMs~\cite{kane}.

In such a case, one expects that these large phases have important
impact on the lightest supersymmetric particle (LSP) relic density
and its detection rates. In Ref.~\cite{shafi,falk} the effect of
SUSY phases on the LSP mass, purity, relic density, elastic cross
section and detection rates has been considered within models with
universal, hence real, gaugino masses. It was shown that the
phases have no significant effect on the LSP relic abundance but a
substantial impact on the detection rates. 
Here, we study the effect of gaugino phases, particularly, we
consider D-brane model recently proposed~\cite{munoz} which is
able to allow large value of phases while the EDM of the neutron
and electron are less than the experimental limit as shown in
Ref.~\cite{kane}. It turns out that the LSP of this model could be
bino or wino like depending on the ratio between $M_1$ and $M_2$.
In the region where the EDMs are smaller than the limit, the mass of the
LSP is very close to the lightest
chargino, hence the co-annhilation between them becomes very
important and it greatly reduces the relic density~\cite{khalil}. The
phases have no important effect on the LSP relic aboundance as in the
case of Ref.~\cite{falk}. However, their effect on the detection
rates is very significant and is larger than what is found in the
case of real gaugino masses~\cite{falk}.

The possibility of non-universal gaugino masses and phases at the
tree level is natural in the type I string theory~\cite{munoz}.
The soft SUSY breaking terms in this class of models depend on the
embedding of the standard model (SM) gauge group in the D-brane
sector. In case of the SM gauge group is not associated with a
single set of branes the gaugino masses are non universal. We
assume that the gauge group $SU(3)_C \times U(1)_Y$ is associated
with one set of five branes (say $5_1$) and $SU(2)_L$ is
associated with a second set $5_2$ . The soft SUSY breaking terms
take the following form~\cite{munoz}.
\begin{eqnarray}
M_1 &= & \sqrt{3} m_{3/2} \cos \theta \Theta_1 e^{-i \alpha_1} =
M_3 = - A ,\\ M_2 &=& \sqrt{3} m_{3/2} \cos \theta \Theta_2 e^{-i
\alpha_2} ,
\end{eqnarray}
where $A$ is the trilinear coupling. The soft scalar mass squareds
are gives by
\begin{eqnarray}
m_Q^2 &=& m^2_L = m_{H_1}^2 = m_{H_2}^2 = m_{3/2} (1-3/2 \sin^2
\theta) ,\\ m_D^2 &=& m^2_U = m_E^2 = m_{3/2} (1-3 \cos^2 \theta),
\end{eqnarray}
and $\Theta_1^2 + \Theta_2^2 = 0$. In this case, by using the
appropriate field redefinitions and the $R$-rotation we end up
with four physical phases, which can not be rotated away. These
phases can be chosen to be: the phase of $M_1$ ($\phi_1$), the
phase of $M_3$ ($\phi_3$), the phase of $A$ ($\phi_A$) and the
phase of $\mu$ ($\phi_{\mu}$). The phase of $B$ is fixed by the
condition that $B \mu$ is real.

The effect of these phases on the EDM of the electron and the
neutron, taking into account the cancellation mechanism between
the different contributions, has been examined in
Ref.~\cite{kane}\footnote{See also Ref.~\cite{arnowitt}}. It was
shown that large values of these phases can be accommodated and
the electron and neutron EDM satisfy the experimental constraint.
It is worthy noticed that the EDM impose a constraint on the ratio
$M_1/M_2$. In fact, to have an overlap between the electron and
neutron EDM allowed regions, $M_2$ should be less than $M_1$, and
as explained in Ref.~\cite{kane}, a precise overlap between these
two regions occurs at $\Theta_1 = 0.85$. Such constraint has an
important impact on the LSP. In this case, we have $M_2$ is the
lightest gaugino at GUT scale. However, at the electroweak (EW)
scale, it turns out that the lightest neutralino is a bino like.
Furthermore, the LSP mass is close to the lightest chargino mass
which is equal to the mass of the next lightest neutralino
$(\tilde{\chi}_2^0)$. Therefore, the co-annhilation between the
bino and the chargino as well as the next to lightest neutralino
are very important and have to be included in the calculation of
the relic density.

We study the effect of the SUSY CP violating phases and the
co-annihilation on the relic density and also on the upper bound
of the LSP mass. Since the LSP is bino like, the annihilation is
predominantly, as usual, into leptons by the exchange of the right
slepton. Without co-annihilation, the constraint on the relic
density $0.025 < \Omega_{LSP} h^2 < 0.22 $ impose sever constraint
on the LSP mass, namely $m_{\chi} < 150$ GeV, and the SUSY phases
have no any significant effect in relaxing such sever constraint
as found in Ref.~\cite{shafi}. Including the co-annihilation of $\chi$
with $\chi_1^+$ and $\tilde{\chi}_2^0$ is very important to reduce the LSP
relic density to an acceptable level.

Given that the LSP is almost pure bino, the co-annihilation
processes are predominantly into fermions. However, since the
coupling of $\tilde{\chi}_2^0-f- \tilde{f}$ is proportional to
$Z_{2j}$, it is smaller than the coupling of $\tilde{\chi}_1^+-f-
\tilde{f'}$. We found that the dominant contribution is due to the
co-annhiliation channel $\tilde{\chi}_1^+ \chi \rightarrow f
\bar{f}$. We also include $\tilde{\chi}_1^+ \chi \rightarrow W^+
\gamma$ channel, estimated to contribute with a few cent. Then, we can
calculate the relic aboundance using the
standard procedure~\cite{grist}. In Fig.1 we show the values of
the LSP relic abundance $\Omega_{\chi} h^2$, estimated with
including the co-annihilations, corresponding to the LSP mass.
\begin{figure}[h]
\psfig{figure=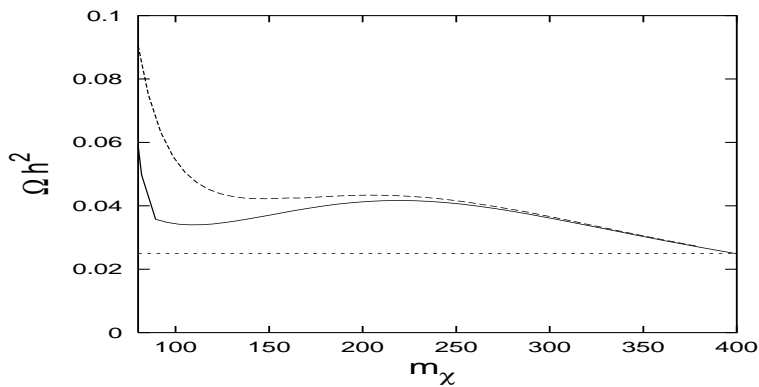,height=5cm,width=10cm} \caption{The LSP
relic abundance with co-annihilation versus its mass, solid line
corresponds to non vanishing phases while the dashed lines
correspnd to vanishing phases.}
\vskip -0.25cm
\end{figure}
This figure shows that the co-annihilation processs have very
significat rule in reducing the values of $\Omega_{\chi} h^2$,
even now we obtain an upper bound on the mass of the LSP from the
lower bound of the relic density, $\Omega_{\chi} h^2 > 0.025$
which leads to $m_{\chi} < 400$ GeV.  Here, also the effect of the
SUSY phases is insignificant and the same upper bound of the LSP
mass is obtained for vanishing and non vanishing phases. It is
important to notice that the gaugino phases especially the phase
of $M_3$ have important impact on having large $\phi_A$ at the EW
scale. It dominantly contributes to the phase of $A$-term during
the renormalization from the GUT scale to EW scale. Thus, the
radiative corrections to $\phi_A$ is very small and the phase of
$A$ is kept large at EW. However, as we have shown, such large
phases are not effecting for the LSP mass and the relic abundance.
In fact, this result is due to two facts, first the LSP is bino so
it slightly depends on the phase of $\mu$, second, the phases are
important if there is a significant mixing in the sfermion mass
matrix. In theses class of models we consider the off diagonal
element are much smaller than the diagonal element. 

As shown in Ref.~\cite{shafi}, the SUSY phases is found to have a
significant effect on the direct detection rate ($R$) and indirect
detection rate ($\Gamma$).  The phase of $\phi_A$ increases the
values of $R$ and $\Gamma$ . Furthermore, the enhancement of the
ratios of the rates with non vanishing $\phi_A$ to the  rates in
the absence of this phase are even large than what is found in
Ref.~\cite{shafi}, since as we explained, here $\phi_A$ has larger
values at EW scale due to the gluino contribution through the
renormalization.
\vskip 0.25cm
\hspace{-0.7cm}This work is supported by a Ministerio de Educacion y
Cultura research grant.

\end{document}